\documentclass[hep,a4paper,fleqn,11pt]{w-hep}
\usepackage{times,cite,w-greek,amsmath,latexsym,amssymb}
\usepackage[]{graphicx}

\newcommand\astp[3]{{\sl Astropart.\ Phys.\ }{\bf #1}, #3 (#2)}

\newcommand\jcap[3]{{\sl J.\ Cosmol.\ Astropart.\ Phys.\ }{\bf #1}, #3 (#2)}
\newcommand\npb[3]{{\sl Nucl.\ Phys.\ }{\bf B#1}, #3 (#2)}

\newcommand\plb[3]{{\sl Phys.\ Lett.\ }{B \bf #1}, #3 (#2)}
\newcommand\prd[3]{{\sl Phys.\ Rev.\ }{D \bf #1}, #3 (#2)}

\newcommand\prl[3]{{\sl Phys.\ Rev.\ Lett.\ }{\bf #1}, #3 (#2)}

\newcommand{\arxiv}[1]{{\tt arXiv:#1}}

\newcommand{\ftn}{\footnotesize}

\newcommand{\ssz}{\scriptsize}

\newcommand{\TeV}{{\mbox{\rm TeV}}}

\newcommand{\GeV}{{\mbox{\rm GeV}}}

\newcommand{\vH}{{\ensuremath{\bar H}}}

\newcommand{\vHi}{{\ensuremath{\bar H_{{\rm I}}}}}
\newcommand{\vrho}{{\ensuremath{\bar\rho}}}

\newcommand{\vq}{{\ensuremath{\bar q}}}
\newcommand{\vqi}{{\ensuremath{\bar q_{\rm I}}}}

\newcommand{\vTi}{{\ensuremath{T_{{\rm I}}}}}
\newcommand{\vti}{{\ensuremath{\vtau_{{\rm I}}}}}

\newcommand{\vtns}{{\ensuremath{\vtau_{{\rm BBN}}}}}
\newcommand{\vtkr}{{\ensuremath{\vtau_{{\rm KR}}}}}
\newcommand{\vtp}{{\ensuremath{\vtau_{{\rm ext}}}}}

\def\openep{\leavevmode\hbox{\normalsize$\iota$\kern-3.8pt$^$-}}
\newcommand{\vtauf}{\ensuremath{\uptau}}
\newcommand{\vtau}{\ensuremath{\uptau}}

\newcommand{\vtf}{{\ensuremath{\vtau_{{\rm F}}}}}

\newcommand{\eem}{\end{matrix}}
\newcommand{\bem}{\begin{matrix}}
\def\beq{\begin{equation}}
\def\eeq{\end{equation}}
\def\bea{\begin{eqnarray}}
\def\eea{\end{eqnarray}}

\newcommand{\Ti}{\ensuremath{T_{\rm I}}}
\newcommand{\Tkr}{\ensuremath{T_{\rm KR}}}

\newcommand{\Omx}{\ensuremath{\Omega_\chi h^2}}
\newcommand{\Domx}{\ensuremath{\Delta\Omega_{\chi}}}
\newcommand{\Omxsc}{\ensuremath{\left.\Omx\right|_{\rm SC}}}

\newcommand\sigv{\ensuremath{\langle \sigma v\rangle}}

\newcommand{\ps}{\ensuremath{e^+}}
\newcommand{\el}{\ensuremath{e^-}}
\newcommand{\mchi}{\ensuremath{\left.\upchi^2\right|_{\rm min}}}

\newcommand{\nt}{\ensuremath{\chi}}

\newcommand{\mx}{{\mbox{$m_\chi$}}}

\newcommand{\Eref}[1]{Eq.~(\ref{#1})}

\newcommand{\cref}[1]{Ref.~\cite{#1}}
\newcommand{\etal}{{\it et al.\/}}
\newcommand{\dof}{{\mbox{\rm d.o.f}}}

\begin{document}



\title[Tracking Quintessence,
WIMP Relic Density, PAMELA and Fermi LAT]{Tracking Quintessence,
WIMP Relic Density,\\ PAMELA and Fermi LAT}

\author[C. Pallis]{C. Pallis}
\address[]{\sl Department of Physics,  University of Patras, GR-265 00
Patras, GREECE}

\begin{abstract}
The generation of an early \emph{kination dominated} (KD) era
within a tracking quintessential model is investigated, the relic
density of the \emph{Weakly Interacting Massive Particles} (WIMPs)
is calculated and we show that it can be enhanced \emph{with
respect to} (w.r.t) its value in the \emph{Standard Cosmology}
(SC). By adjusting the parameters of the quintessential scenario,
the cold dark matter abundance in the universe can become
compatible with large values for the annihilation cross section
times the velocity of the WIMPs. Using these values and assuming
that the WIMPs annihilate predominantly to $\mu^+\mu^-$, we
calculate the induced fluxes of $e^\pm$ cosmic rays and fit the
current PAMELA and Fermi-LAT data. We achieve rather good fits in
conjunction with a marginal fulfillment of the restriction arisen
from the \emph{Cosmic Microwave Background} (CMB).
\end{abstract}

\maketitle \thispagestyle{empty}

\setcounter{page}{1}

\section{Introduction}

A plethora of recent data \cite{wmap} indicates that the two major
components of the universe are the \emph{Cold Dark Matter} (CDM)
and \emph{Dark Energy} (DE). The DE component can be explained
with the introduction of a slowly evolving today scalar field,
$q$, called quintessence whereas WIMPs, $\chi$, are the most
natural candidates to account for CDM. In this talk, which is
based on Ref.~\cite{patra}, we reconsider (Sec.~\ref{sec1}) the
creation of an early era dominated by the kinetic energy of $q$ in
the context of a tracking quintessential model \cite{mas}. We show
(Sec.~\ref{sec2}) that if $\chi$ decouples from the cosmic fluid
during this era, its relic abundance, $\Omx$ can be significantly
enhanced w.r.t. its value in SC. This enhancement of $\Omx$
assists us to interpret, through the $\chi$'s annihilation in the
galaxy (Sec.~\ref{sec3}), the reported \cite{pamela, Fermi} excess
on the positron ($e^+$) and/or electron ($\el$) \emph{cosmic-ray}
(CR) flux under the assumption that $\chi$'s annihilate
predominantly into $\mu^+\mu^-$ (Sec.~\ref{sec4}). Throughout, the
subscript $0$ [I] is referred to present-day values [to values at
the onset of our scenario] and $\bar \rho_{i}=\rho_{i}/\rho_{\rm
c0}~(i=q,\mbox{R and M})$ where $\rho_{\rm
c0}=8.1\cdot10^{-47}h^2~{\rm GeV^4}$ with $h=0.72$.

\section{The Tracking Quintessential Model}\label{sec1}

The quintessence field, $q$, of our \emph{quintessential scenario}
(QS) satisfies the equation:
\beq \ddot q+3H\dot
q+dV/dq=0,~~\mbox{with}\hspace*{0.15cm}V={M^{4+a}/
q^a}+b\,H^2q^2/2\hspace*{0.15cm}\mbox{and}\hspace*{0.15cm}H/H_0=\vH=\sqrt{\vrho_q
+\vrho_{\rm R}+\vrho_{\rm M}}.~~~~~\label{qeq} \eeq
Here $H$ is the Hubble parameter, dot denotes derivative w.r.t the
cosmic time $t$, $\rho_q=\dot q^2/2+V$, $\rho_{{\rm
R}}\simeq\rho_{\rm R0}\exp{(-4\vtau)}$ and $\rho_{\rm M}=\rho_{\rm
M0}\exp{(-3\vtau)}$ is the $q$, radiation and matter energy
density respectively, $\vtau=\ln(R/R_0)$ is the logarithmic time
and $R$ is the scale factor of the universe.

We impose on our QS the following constraints:

\begin{enumerate}
\item[{\ftn\sf(a)}] \label{domk} \emph{Initial Domination of
Kination.} We focus our attention in the range of parameters with
$\Omega_{q\rm I}=\Omega_q(\vti)=1$ where
$\Omega_q\simeq\rho_q/(\rho_q+\rho_{{\rm R}}+\rho_{\rm M})$ is the
quintessential energy-density parameter.

\item[{\ftn\sf(b)}] \emph{Inflationary Constraint.} Assuming that
the power spectrum of the curvature perturbations is generated by
an early inflationary stage, we impose the bound
$\vHi\lesssim1.72\cdot10^{56}$.

\item[{\ftn\sf(c)}]\label{para} \emph{Nucleosynthesis (BBN)
Constraint.} At the onset of BBN, $\vtns=-22.3$, $\rho_q$ is to be
sufficiently suppressed compared to $\rho_{{\rm R}}$, i.e.,
\cite{oliven} $\Omega_q(\vtns)\leq0.21$ at $95\%$ \emph{confidence
level} (c.l.).

\item[{\ftn\sf (d)}]\label{rhoq0} \emph{DE and Cosmic Coincidence
Constraint.} These two requirements can be addressed if we demand
\cite{mas} $\Omega_{q0}=\vrho_{q0}=0.74$ and
$d^2V(\vtau=0)/dq^2\simeq H^2_0$.

\item[{\ftn\sf(e)}]\label{wq} \emph{Acceleration Constraint.}
Successful quintessence has to account for the present-day
acceleration of the universe, i.e. \cite{wmap} $-1.12\leq
w_q(0)\leq-0.86~~\mbox{($95\%$ c.l.)}$, where $w_q=(\dot
q^2/2-V)/(\dot q^2/2+V)$ is the barotropic index of the $q$-field.

\end{enumerate}

\begin{figure}[t]\vspace*{-0.6cm}
\begin{minipage}{75mm}
\includegraphics[height=2.7in,angle=-90]{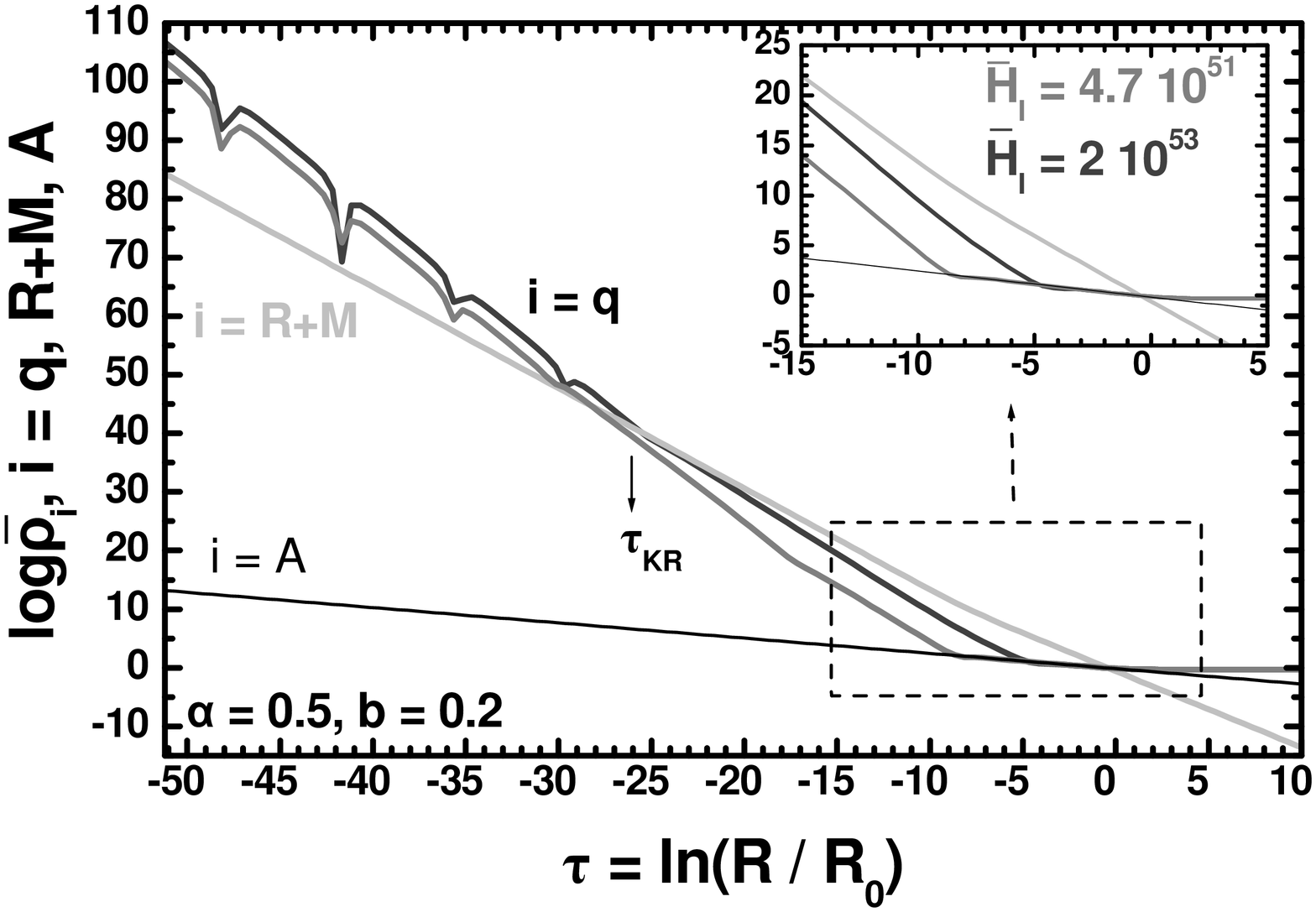}
\end{minipage}
\hfil \hspace*{3mm}\begin{minipage}{70mm} \caption{The evolution
of $\log\vrho_i$ with $i=q$ (gray [dark gray] line), R+M (light
gray line) and A (thick line) as a function of $\vtau$ for
$\vqi=0.01$, $a=0.5$, $b=0.2$, $\Ti=10^9~\GeV$, $M=4.8~{\rm eV}$
and $\vHi=4.7\cdot10^{51}$ [$\vHi=2\cdot10^{53}$] -- note that
$\vrho_{\rm R+M}=\vrho_{\rm R}+\vrho_{\rm M}$. We observe that
although the used $\vHi$'s differ by two orders of magnitude, both
$\vrho_q$'s reach $\vrho_{\rm A}$ highlighting thereby the
insensitivity of our QS to the initial conditions.}\label{fig1}
\end{minipage}

\end{figure}

Solving \Eref{qeq} we find that $q$ undergoes four phases during
its evolution -- see Fig.~\ref{fig1}. For $\vtau<\vtkr$ the
universe and $q$ is dominated by $\dot q/2\gg V$ and therefore we
get a KD era, during which  $q$ is set in anharmonic oscillations
for $b>0$. In particular, $\vq$ develops extrema at
\beq \label{tmax}\vtp\simeq(2k+1)\sqrt{1\over
b}{\pi\over2}+\vti,~~\mbox{with}\hspace*{0.15cm}k=0,1,2,...\eeq
For $\vtau>\vtkr$ the universe becomes initially radiation and
then matter dominated whereas $\rho_q$ is dominated initially by
$\dot q/2$ and then by $V$. As $\vtau$ approaches 0 the system in
\Eref{qeq} admits \cite{mas} a tracking solution since the energy
density of the attractor:
\beq \label{rA} \vrho_{\rm A}\propto\exp\left(-3(1+w_q^{\rm
fp})\vtau\right)~~\mbox{with}\hspace*{0.15cm}w_q^{\rm
fp}=-{2/(a+2)} \eeq
tracks $\vrho_{\rm M}$ until it outstrips and dominates the
current expansion of the universe. It can be shown \cite{mas,
patra} that $b>0$ ensures the coexistence of an early KD phase
with the achievement of the tracking solution in time. Moreover,
for $a<0.6$ the requirement 2-{\ftn \sf (e)} can be marginally
fulfilled, too.

\section{The WIMP Relic Density}\label{sec2}

The relic density of a WIMP $\chi$, $\Omega_{\chi}h^2$, with mass
$m_{\chi}$ is calculated using the formula:
\beq\Omega_{\chi}h^2 = 2.748 \cdot 10^8\ \left({n_{\chi0}/
s_0}\right)\
\left({m_{\chi}/\mbox{GeV}}\right),~~\mbox{where}\hspace*{0.15cm}
\dot n_\chi+3Hn_\chi+\sigv\left(n_\chi^2-n_{\chi}^{\rm
eq2}\right)=0~~~~~~~~\eeq
is the Boltzmann equation which governs the evolution of the
$\chi$'s number density, $n_{\chi}$. Also, $n_{\chi}^{\rm
eq}\propto x^{3/2}\>e^{-1/x}$ with $x=T/m_{\chi}$ is the
equilibrium configuration of $n_{\chi}$, $\sigv$ is the
thermal-averaged cross section times the velocity of $\chi$'s and
$s\propto T^3$ is the entropy density with $T$ the temperature.

The decoupling of $\chi$ from the cosmic fluid during the QS and
SC is visualized in Fig.~2-{\ftn\sf (a)}. We observe that, in both
cases, the current $\rho_{\chi}/s$ follows $\rho^{\rm
eq}_{\chi}/s$ and at some $\vtau=\vtf$, $\rho_{\chi}/s$ dominates
over $\rho^{\rm eq}_{\chi}/s$ and remains constant until today.
For the selected $\mx$ and $\sigv$ we obtain an enhancement of
$\Omx$ within QS w.r.t its value in the SC, $\Omxsc$, since
$\Omx=0.11$ whereas $\Omxsc=0.0045$. This enhancement can be
further analyzed, by defining $\Domx={\Omx/\Omxsc-1}$. The
behavior of $\Domx$ as a function of the free parameters of the QS
can be inferred from Fig.~2-{\ftn \sf (b)}. For $b=0$ we obtain a
pure KD era and $\Domx$ increases when $\mx$ increases or $\sigv$
decreases. For $b\neq0$, $\Domx$ depends crucially on the
hierarchy between $\vtf$ and $\vtp$. As $\mx$ increases above
$0.1~\TeV$, $\vtf$ decreases and moves closer to $\vtp$ and
$\Domx$ decreases with its minimum $\left.\Domx\right|_{\rm min}$
occurring at $\mx=\left.\mx\right|_{\rm min}$ which corresponds to
$\vtau_{\rm F}^{\rm min}\simeq\vtp$.

\begin{figure}\vspace*{-0.6cm}
\includegraphics[height=2.7in,angle=-90]{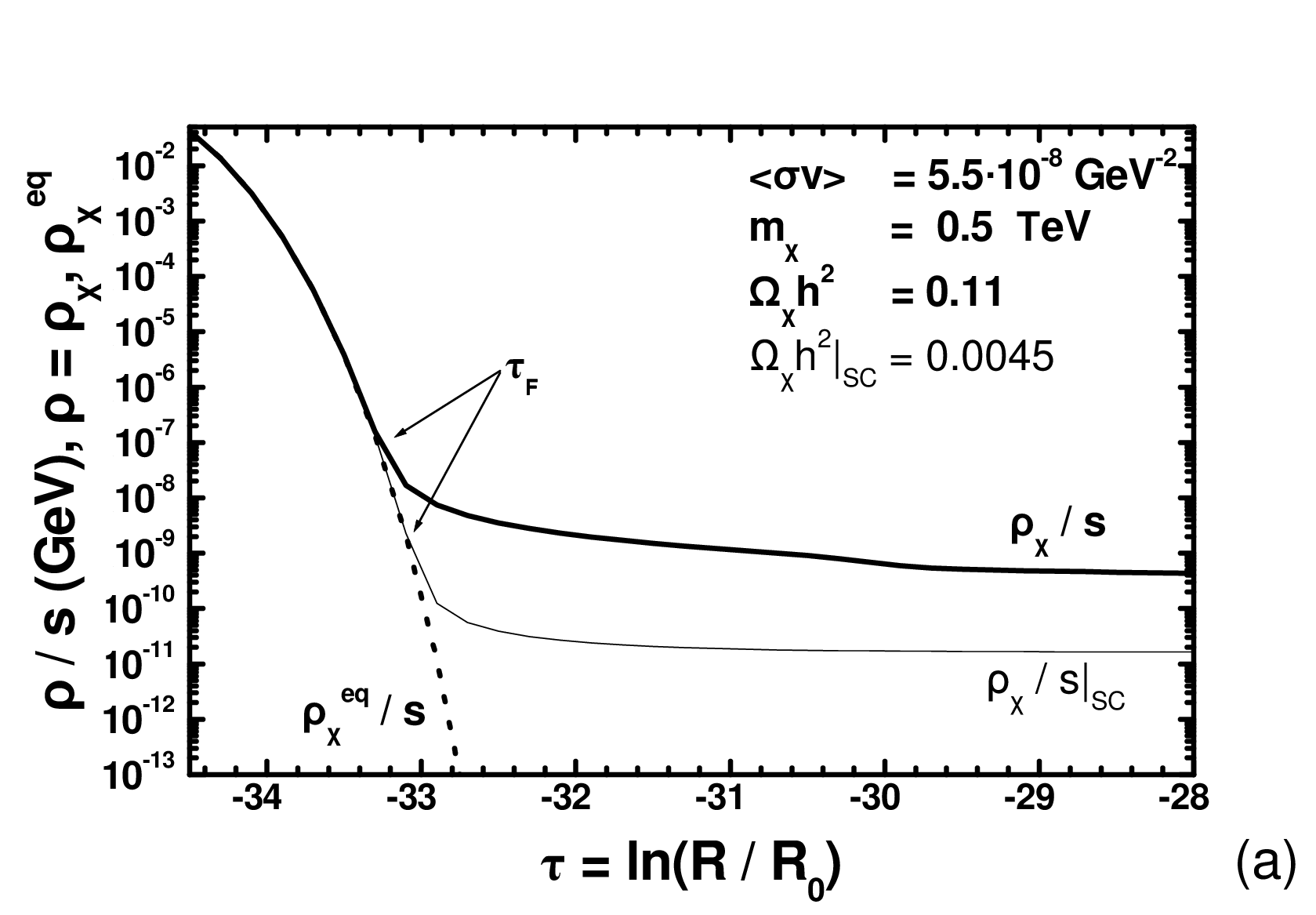}
\hfil
\includegraphics[height=2.7in,angle=-90]{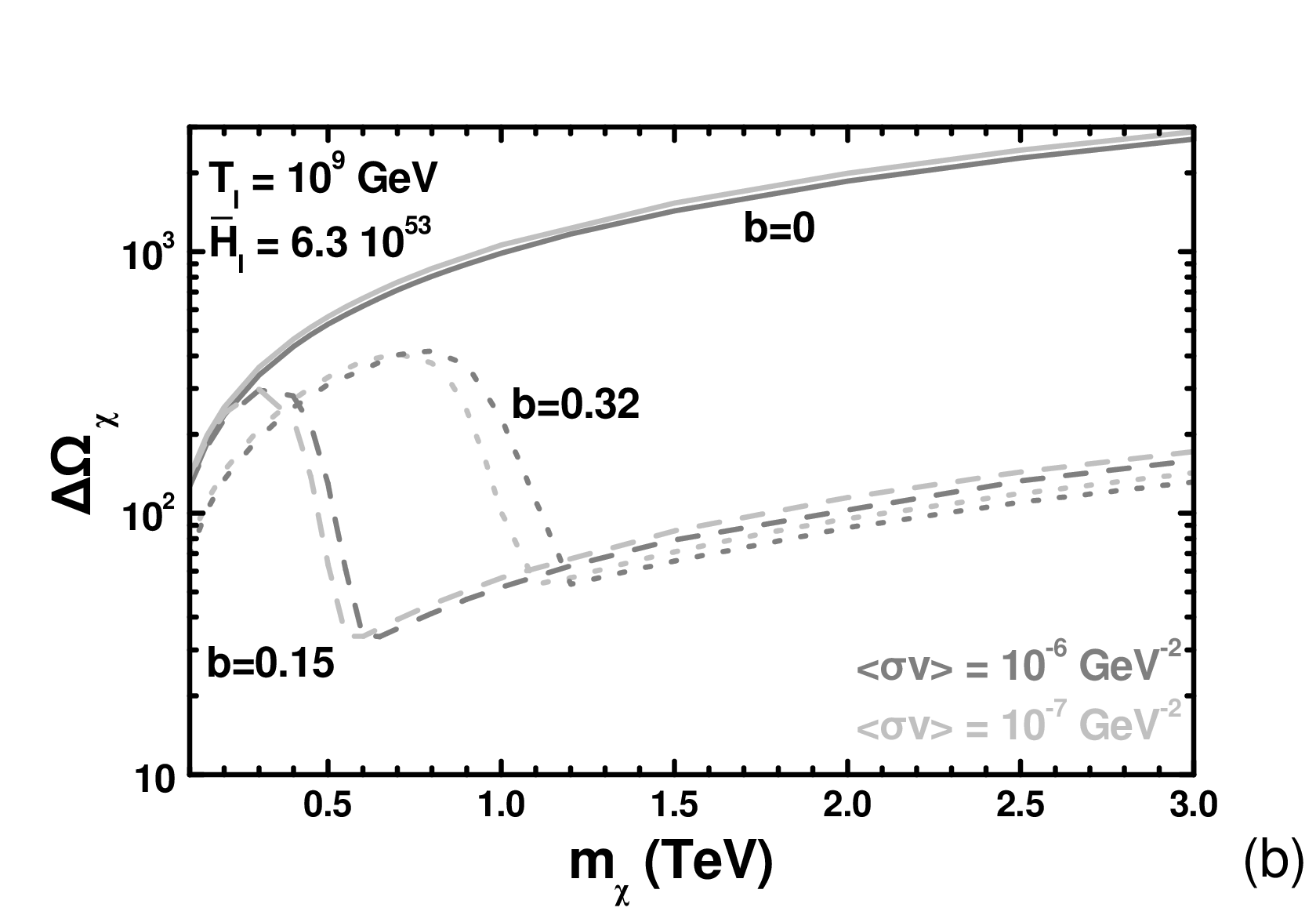}
\caption{{\ssz\sf (a)} The evolution as a function of $\vtauf$ of
the quantities $\rho^{\rm eq}_\chi/s=\mx n^{\rm eq}_{\chi}/s$
(dotted line) and $\rho_\chi/s=\mx n_{\chi}/s$ (thick [thin] solid
lines) for the QS [SC]; {\ssz\sf (b)} $\Domx$ versus $m_\chi$ for
the QS with $a=0.5$, $\vHi=6.3\cdot10^{53}$, $\vTi=10^9~\GeV$,
$\langle\sigma v\rangle=10^{-6}~\GeV^{-2}~[\langle\sigma
v\rangle=10^{-7}~\GeV^{-2}]$ (gray [light gray] lines) and $b=0$
(solid lines), $b=0.15$ (dashed lines) and $b=0.32$ (dotted
lines).}\label{fig2}
\end{figure}

\section{$e^\pm$-CRs From WIMP Annihilation}\label{sec3}

Residual $\chi$'s annihilations in the galaxy induce a $\ps$ flux
per energy at Earth which is given by
\beq \Phi^{\nt\nt}_{e^+} (E) = {1\over2}\frac{v_{e^+}}{4\pi b(E)}
\, \left( \frac{\rho_{\odot} }{m_{\nt}} \right )^2 \, \sigv \;
\int_E^{m_{\nt} }\; d E' \, I\left(\uplambda_{D}(E,E') \right)\,
\frac{d N_{e^+}}{ d E'_{e^+} } \;, ~\mbox{where}~~\label{fp}\eeq
$v_{e^+}$ is the velocity of $\ps$, $\rho_{\odot}=0.3~\GeV/{\rm
cm}^3$ is the local CDM density, $b(E)=E^2/(\GeV\,t_E)$ with $t_E
= 10^{16}~{\rm s}$ is the energy loss rate function and
$dN_{e^+}/dE_{e^+}$ denotes the energy distribution of $\ps$'s per
$\chi$ annihilation and can be found in Ref.~\cite{chi2}. Also,
$I(\lambda_D)$ is the dimensionless halo function which fully
encodes the galactic astrophysics and can be read off from
Ref.~\cite{strumia}

There are three sources of uncertainty in our computation: the CDM
distribution, the propagation of $\chi$ annihilation products and
the astrophysical backgrounds. We adopt {\ftn \sf (a)} the
isothermal halo profile, to avoid troubles with observations on
$\gamma$-CRs; {\ftn \sf (b)} the MED propagation model, which
provides the best fits to the combinations of the two data-sets;
and {\ftn \sf (c)} commonly \cite{strumia} assumed background
$e^\pm$ fluxes normalized with the Fermi-LAT data. Adding the
latter contributions to the one in \Eref{fp} we get the total
fluxes, $\Phi_{e^\pm}$.

In order to qualify our fittings to the experimental data, we
define the $\upchi^2$ variables as follows:
\beq \upchi^2_A =\sum_{i=1}^{N_A} {\left(F_{Ai}^{\rm obs} -
F_{Ai}^{\rm th}\right)^2 \over \left(\Delta F_{Ai}^{\rm
obs}\right)^2 },~~\mbox{with}\hspace*{0.15cm}F_A=\left\{\bem
\Phi_{e^+}/\left(\Phi_{e^+}+\Phi_{e^-}\right) &\hspace*{-0.3cm}
\mbox{and $N_A=7$}& \hspace*{-0.2cm}\mbox{for PAMELA}, \hfill \cr
E^3_{e^+}\left(\Phi_{e^+}+\Phi_{e^-}\right)
&\hspace*{-0.2cm}\mbox{and $N_A= 26$}& \hspace*{-0.2cm}\mbox{for
Fermi LAT} , \hfill \cr\eem
\right.\eeq
where $i$ runs over the data points of each experiment $A$,
``${\rm obs}$'' [``${\rm th}$''] stands for measured
[theoretically predicted] values. The best fits to the combined
experimental data can be achieved with $\mx\simeq1.28~\TeV$ and
$\sigv\simeq1.95\cdot10^{-6}~\GeV^{-2}$ resulting to
$(\upchi_1^2+\upchi_2^2)/\dof=24/31$.

\section{Results}\label{sec4}

To systematize our approach, we can define regions in the
$\mx-\sigv$ plane which are favored at $95\%$ c.l. [$99\%$ c.l.]
by the various experimental data on the $e^\pm$-CRs demanding
$$\upchi^2\lesssim\mchi+6 ~~\Big[\upchi^2\lesssim\mchi+9.2\Big]~~\mbox{with}\hspace*{0.15cm}\upchi^2=\left\{\bem
\upchi^2_1\hfill & \mbox{for PAMELA}, \hfill \cr
\upchi^2_1+\upchi^2_2~\hfill &\mbox{for PAMELA and Fermi LAT},
\hfill \cr
\eem \right.~~$$
where $\mchi$ can be extracted numerically by minimization of
$\upchi^2$ w.r.t $\mx$ and $\sigv$.

The large $\sigv$'s which are required in order to fit the
experimental data on $e^\pm$-CRs are to be consistent with a
number of requirements so as the interpretation of the data on
$e^\pm$-CRs via CDM annihilation in the galaxy is fully
acceptable. All in all, we impose the following constraints:

\begin{itemize}
\item[\ftn\sf (a)] \emph{Constraint from the CDM abundance
\cite{wmap}}: $0.097\lesssim \Omx\lesssim0.12.$
\item[\ftn\sf (b)] \emph{BBN constraint \cite{moroiNS}}:
$\sigv\leq8.6\cdot10^{-5}~\GeV^{-2}\;\left({\mx/\TeV}\right).$
\item[\ftn\sf (c)]  \emph{CMB constraint \cite{CMB}}:
$\sigv\leq1.3\cdot10^{-6}~\GeV^{-2}\;\left({\mx/\TeV}\right).$
\item[\ftn\sf (d)] \emph{Constraint from the $\gamma$-CRs
\cite{gstrumia}}: $\sigv\lesssim 4\cdot10^{-6}~\GeV^{-2}$ for the
isothermal halo profile.
\item[\ftn\sf (e)] \emph{Unitarity constraint}:
$\sigv\leq8\pi~\GeV^{-2}\;\left({\mx/\GeV}\right)^{-2}.$

\end{itemize}

\begin{figure}[t]\vspace*{-0.6cm}
\begin{minipage}{75mm}
\includegraphics[height=2.7in,angle=-90]{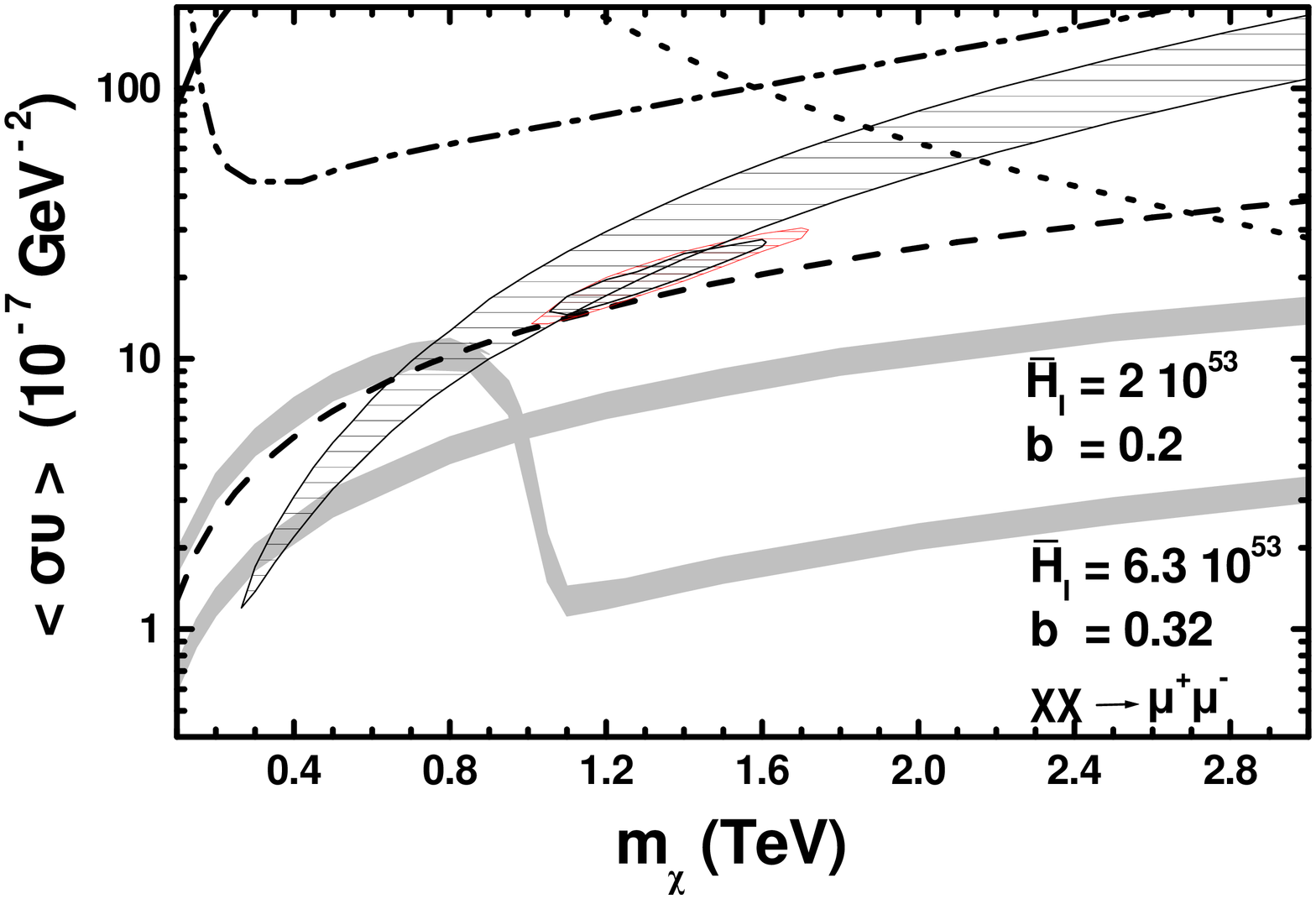}

\label{fig:8}
\end{minipage}
\hfil
\begin{minipage}{65mm}\renewcommand{\arraystretch}{1.1}
\begin{tabular}{@{}cccc@{}}
\hline
$\mx$&\multicolumn{3}{c}{$1.12~~\TeV$}\\
$\sigv$&\multicolumn{3}{c}{$1.44\cdot10^{-6}~~\GeV^{-2}$}\\
\hline
$\left.\Omx\right|_{\rm SC}$&\multicolumn{3}{c}{$0.00019$}\\\hline
\multicolumn{4}{c}{\sc Parameters Yielding $\Omx=0.11$ }\\
\multicolumn{4}{c}{\sc in our QS $\left(a=0.5,~\vTi=10^9~\GeV\right)$}\\
\hline
$b$& $0$& $0.08$&$0.18$\\
$\vHi/10^{53}$ & $3.1$ & $3.4$& $4.7$\\\hline
$\Tkr~(\GeV)$ & $0.019$ & $0.005$& $0.009$\\ \hline
\end{tabular}
\end{minipage}
\caption{Restrictions in the $\mx-\sigv$ plane for
$a=0.5,~\vTi=10^9~\GeV$ and several $b$'s and $\vHi$'s indicated
in the graph. The light gray shaded areas are allowed the
constraint 5-{\ssz\sf (a)}, the sparse black hatched area is
preferred at $95\%$ c.l. by the PAMELA data and the dense black
[red] hatched areas are preferred at $95\%$ c.l. [$99\%$ c.l.] by
the PAMELA and Fermi-LAT data. Regions above the black solid,
dashed, dot-dashed and dotted lines are ruled out by the upper
bounds on $\sigv$ from the constraints 5-{\ssz\sf (b)}, 5-{\ssz\sf
(c)} 5-{\ssz\sf (d)} and 5-{\ssz\sf (e)}, respectively. The
best-fit $\left(\mx, \sigv\right)$ for the combination of PAMELA
and Fermi-LAT data which is consistent with all the constraints is
given in the table. Shown are also $\left.\Omx\right|_{\rm SC}$
and several $b$'s and $\vHi$'s (and the resulting $\Tkr$'s)
leading to $\Omx\simeq0.11$ in our QS.}\label{fig3}
\end{figure}

Imposing all the constraints above we can delineate our findings
in the $\mx-\sigv$ plane as in Fig.~\ref{fig3}. A simultaneous
interpretation of the $e^\pm$-CR anomalies consistently with the
various constraints can be achieved in the regions where the gray
shaded areas overlap the lined ones below the dashed lines. We
observe that part of the region favored at $99\%$ c.l. by PAMELA
and Fermi LAT is allowed. The best-fit $\left(\mx, \sigv\right)$
-- with $\upchi^2/\dof = 33/31$ -- which saturates the most
stringent (CMB) bound is arranged in the Table of Fig.~\ref{fig3}.
We remark that the requirement 5-{\sf\ftn (a)} is violated within
SC but can be met by adjusting the parameters of the QKS. In all
cases we obtain $\Tkr<0.02~\GeV$ with $\Tkr$ being the transition
temperature to the conventional RD era. It remains the
construction of a particle model with the appropriate couplings so
that $\chi$'s annihilate into $\mu^+\mu^-$ with the desired
$\sigv$ derived self-consistently with the (s)particle spectrum.

\renewcommand{\arraystretch}{1.}


\begin{acknowledgement}
\hspace*{-0.25cm} This research was funded by the FP6 Marie Curie
Excellence grant MEXT-CT-2004-014297.
\end{acknowledgement}

\def\bstname{fdp}

\end{document}